\documentclass[a4paper,12pt]{article}
\usepackage{amssymb}
\usepackage{amsmath}
\usepackage{epsfig}
\usepackage{stmaryrd}
\usepackage{graphics}

\usepackage{latexsym}
\usepackage{rotating}
\title
{Semi-direct sums of Lie algebras and \\discrete integrable
couplings
 }
\author {
Wen-Xiu Ma\thanks{Email: mawx@math.usf.edu (W.X. Ma), Tel:
(813)974-9563, Fax: (813)974-2700}
\\{\small Department of Mathematics, University of South Florida, Tampa, FL 33620-5700,
USA}\vspace{2mm}
\\{Xi-Xiang Xu\thanks{Email: xu$\_$xixiang@sohu.com (X.X. Xu)} }
\\{\small College of Science, Shandong University of Science and Technology, Qingdao 266510, P.R. China}\vspace{2mm}
\\{Yufeng Zhang\thanks{Email: zhang$\_$yfshandong@163.com (Y. Zhang)} }
\\{\small School of Mathematics, Liaoning Normal University, Dalian 116029, P.R. China}
}  \date{\nonumber}
\begin{document}
\maketitle
\date{\nonumber}

\newcommand{\R}{\mathbb{R}}

\numberwithin{equation}{section}

\begin{abstract}
A relation between semi-direct sums of Lie algebras and integrable
couplings of lattice equations is established, and a practicable
way to construct integrable couplings is further proposed. An
application of the resulting general theory to the generalized
Toda spectral problem yields two classes of integrable couplings
for the generalized Toda hierarchy of lattice equations. The
construction of integrable couplings using semi-direct sums of Lie
algebras provides a good source of information on complete
classification of integrable lattice equations.


\end{abstract}

\newtheorem{thm}{Theorem}[section]
\newtheorem{Le}{Lemma}[section]

\setlength{\baselineskip}{17.8pt}
\def \part {\partial}
\def \be {\begin{equation}}
\def \ee {\end{equation}}
\def \bea {\begin{eqnarray}}
\def \eea {\end{eqnarray}}
\def \ba {\begin{array}}
\def \ea {\end{array}}
\def \si {\sigma}
\def \al {\alpha}
\def \la {\lambda}
\def \D {\displaystyle}

\section{Introduction}

Integrable couplings have been receiving growing attention
recently. A few ways to construct integrable couplings are
presented by using perturbations
\cite{MaF-CSF1996,Ma-MAA2000,MaF-PLA1996}, enlarging spectral
problems \cite{Ma-PLA2003,Ma-JMP2005}, and creating new loop Lie
algebras
%also introduced various loop Lie algebras for generating integrable couplings
\cite{GuoZ-JMP2003,Zhang-CSF2004}.

The problem of integrable couplings can be expressed as follows
\cite{Ma-MAA2000}:
  {\it For a
given integrable system, how can we construct a non-trivial system
of differential equations which is still integrable and includes
the original integrable system as a sub-system?} Obviously, a
change of orders of equations in a system does not lose
integrability of the system. Therefore, up to a permutation, an
integrable coupling of a given integrable system $u_t=K(u)$ is
given by a bigger and triangular system:
\begin{equation} \left. \begin{array} {l} u_t=K(u),\
v_t=S(u,v).
\end{array} \right. \nonumber
\end{equation}
 The vector-valued function $S$ should satisfy the non-triviality
condition $
\part S /\part
[u] \ne 0$, where $[u]=(u,D_xu,D_x^2u,\cdots)$ and $D_x^nu$
denotes a vector consisting of all derivatives of $u$ of order $n$
with respect to the space variable $x$. The above non-triviality
condition means that the second sub-system involves the dependent
variables of the first sub-system (i.e. the original system), and
thus it guarantees that trivial diagonal systems with $S(u,v)=
S(v)$ are not within our discussion.

A basic integrable coupling of an integrable system $u_t=K(u)$ is
given by
\begin{equation} u_t=K(u),\ v_t=K'(u)[v],
\label{eq:basicic:ma142}
\end{equation}
which can be generated by a perturbation around a solution of the
system $u_t=K(u)$ \cite{MaF-CSF1996}. In the above system and
elsewhere throughout this paper, $P'(u)[v]$ denotes the Gateaux
derivative of $P(u)\equiv P(u,D_xu,\cdots) $ with respect to $u$
in a direction $v$, i.e.,
\[ P'(u)[v]=\frac {\partial }{\partial
\varepsilon}P(u+\varepsilon v) \Bigr.\Bigl|_{\varepsilon =0}=
\frac {\partial }{\partial \varepsilon}P(u+\varepsilon
v,D_xu+\varepsilon D_xv,\cdots) \Bigr.\Bigl|_{\varepsilon =0} .\]
Obviously, the second sub-system $v_t=K'(u)[v]$ in the above
integrable coupling (\ref{eq:basicic:ma142}) is linear with
respect to $v$. Moreover, a symmetry $S(u)$ of the system
$u_t=K(u)$ leads to a solution $(u,S(u))$ to the integrable
coupling (\ref{eq:basicic:ma142}). However, the second component
$v$ of a solution $(u,v)$ to the integrable coupling
(\ref{eq:basicic:ma142}) is generally not a symmetry of the system
$u_t=K(u)$. This is because $v$ satisfies the linearized system
$v_t=K'(u)[v]$ only for one solution, not for all solutions of the
system $u_t=K(u)$. Therefore, the simple integrable coupling
(\ref{eq:basicic:ma142}) is already a generalization of the
symmetry problem. Another basic integrable coupling of an
integrable system $u_t=K(u)$ reads as
\begin{equation}
u_t=K(u),\ v_t=K'(u)[v]+K(u).\label{eq:2ndbasicic:ma142}
\end{equation}
This system has a set of hereditary recursion operators
\cite{Ma-MAA2000}
\begin{equation}
\Phi(\beta_1,\beta _2)=\left(\ba {cc} \beta _1\Phi (u) & 0\vspace{2mm}\\
\beta _1\Phi '(u)[v]+\beta _2 \Phi (u) &\beta _1\Phi (u)  \ea
\right) \nonumber
\end{equation}
with two arbitrary constants $\beta_1$ and $\beta _2$, if the
original system $u_t=K(u)$ has a hereditary recursion operator
$\Phi(u)$. Therefore, integrable couplings possess richer
integrable structures than the original integrable systems.

The study of integrable couplings provides clues towards complete
classification of integrable systems. Let us first observe
classification of square matrix spectral problems through the
Jordan blocks under similar transformations of matrices. Each
triangular Jordan block corresponds to an undecomposable
sub-system in a given integrable system. Now, note that an
arbitrary Lie algebra has a semi-direct sum structure of a
solvable Lie algebra and a semi-simple Lie algebra
\cite{FrappatSS-book2000}, and we will see that semi-direct sums
of Lie algebras can result in integrable couplings. These imply
that the study of integrable couplings through semi-direct sums of
Lie algebras is an inevitable step towards complete classification
of integrable systems with an arbitrary number of components, from
a point of view of Lie algebras.

The study of integrable couplings also generates interesting
mathematical structures such as Lax pairs with several spectral
parameters \cite{Sakovich-JNMP1998,Sakovich-JNMP1999,Ma-MAA2000},
integrable constrained flows with higher multiplicity
\cite{MaZ-PA2001,MaZ-CSF2002}, local bi-Hamiltonian structures in
higher dimensions \cite{Ma-JMP2002} and hereditary recursion
operators of higher order \cite{Ma-MAA2000,FanZ-CSF2005}. Very
recently, we have proposed a relation between semi-direct sums of
Lie algebras and integrable couplings of continuous soliton
equations, which provides an interesting and systematic approach
to integrable couplings of continuous soliton equations
\cite{MaXZ-PLA2006}. In this paper, we would like to discuss the
problem of discrete integrable couplings and develop a theory for
constructing discrete integrable couplings by use of semi-direct
sums of Lie algebras.

Throughout our discussion, we denote by $E$ the shift operator,
write
\begin{equation}
(E^mx)(n)=x^{(m)}(n)=x(m+n),\ \textrm{where}\  x:\mathbb{Z}\to
\mathbb{R},\ m,n\in \mathbb{Z},
 \label{eq:defofE:pma-xuzhang200509}
\end{equation}
and adopt an inverse of the difference operator $E-1$ as follows
\begin{equation}
(E-1)^{-1}= \frac 12 \bigl( \sum_{k=-\infty}^{-1}
-\sum_{k=0}^\infty \bigr)E^k.\label{eq:ivofe-1:pma-xuzhang200509}
\end{equation}
 Let $G$ be a matrix
Lie algebra with the standard Lie bracket $[A,B]=AB-BA$, and
closed under matrix multiplication: $AB\in G$ for all $A,B\in G$.
We assume that an integrable lattice equation (or system) of
evolution type
\begin{equation} u_t=K(u)=K(u, Eu,E^{-1}u,\cdots)\label{eq:ile:pma-xuzhang200509}\end{equation}
 is associated with $G$, where $u=u(n,t)$ is a dependent variable.
  More precisely, there is a pair of square matrices $U$ and $V$
in $G$, called a Lax pair, so that the discrete spatial matrix
spectral problem
\begin{equation} E\phi=U\phi=U(u,\lambda )\phi
\label{eq:sp:pma-xuzhang200509}
\end{equation}
and the associated discrete temporal matrix spectral problem
\begin{equation}
\phi_{t}=V\phi =V(u,Eu,E^{-1}u,\cdots;\lambda )\phi
,\label{eq:spfort:pma-xuzhang200509}
\end{equation}
where $\lambda $ is a spectral parameter and $\phi$ is an
eigenfunction, generate \cite{Tu-JPA1990,MaF-JMP1999} the
integrable lattice equation (\ref{eq:ile:pma-xuzhang200509})
through their isospectral (i.e., $\lambda _t=0$) compatibility
condition
\begin{equation}
U_t=(EV)U-UV,\label{eq:dzce:pma-xuzhang200509}
\end{equation}
which is called a discrete zero curvature equation.
 In other words,
we have \begin{equation} U'(u)[K]=(EV)U-UV,
\label{eq:algeqnofcompcon:pma-xuzhang200509}
\end{equation}
where $U'(u)[K]$ denotes the Gateaux derivative as above.
 In a non-isospectral case, e.g., $\lambda_t=f(\lambda )$,
then we have
\begin{equation}
U'(u)[K]+fU_{\lambda }=(EV)U-UV,
\label{eq:algeqnofnonisospectralcompcon:pma-xuzhang200509}
\end{equation}
 where $U_\lambda $ is
a partial derivative of $U$ with respect to $\lambda$. Based on
(\ref{eq:sp:pma-xuzhang200509}) and
(\ref{eq:spfort:pma-xuzhang200509}), the lattice equation
(\ref{eq:ile:pma-xuzhang200509}) can often be solved by the
inverse scattering transform (for example, see
\cite{AblowitzL-JMP1975}). There are also a few interesting Lie
algebraic structures hidden behind the equation
(\ref{eq:algeqnofnonisospectralcompcon:pma-xuzhang200509}) (see
\cite{MaF-JMP1999} for more information). An integrable hierarchy
and its master symmetry hierarchy usually correspond to the
isospectral case and the non-isospectral case $\lambda
_t=\lambda^n,\ n\ge 0$, respectively. These two hierarchies
constitute a semi-direct sum of Lie algebras, each of which
consists of symmetries in one hierarchy. The spatial matrix
spectral problem (\ref{eq:sp:pma-xuzhang200509}) is our starting
point in constructing discrete integrable couplings. The closure
property of the Lie algebra $G$ under matrix multiplication
guarantees that $(EV)U-UV$ is still in $G$ so that the discrete
zero curvature equation (\ref{eq:dzce:pma-xuzhang200509}) makes
sense.

In what follows, we are going to establish a relation between
semi-direct sums of Lie algebras and integrable couplings of
lattice equations and a technically-practicable way to generate
integrable couplings through semi-direct sums of Lie algebras. The
resulting general theory will be used to generate two classes of
integrable couplings for the generalized Toda hierarchy presented
in \cite{MaX-JPA2004}. It will also be indicated that the
construction of integrable couplings using semi-direct sums of Lie
algebras provides a good source of information about
classification of integrable lattice equations. A few concluding
remarks will be given in the last section.

\section{Constructing integrable couplings using semi-direct sums of Lie algebras}

\subsection{Generating scheme}

Assume that the lattice equation (\ref{eq:ile:pma-xuzhang200509})
has a Lax pair $(U,V) $ in a matrix Lie algebra $G$ closed under
matrix multiplication.

To construct an integrable coupling of the lattice equation
(\ref{eq:ile:pma-xuzhang200509}), we use semi-direct sums of Lie
algebras to enlarge the original Lie algebra $G$. Take another
matrix Lie algebra $G_c$ closed under matrix multiplication and
then form a semi-direct sum $\bar G$ of $G$ and $G_c$:
\begin{equation} \bar G =G\inplus G_c. \end{equation}
The notion of semi-direct sums means that $G$ and $G_c$ satisfy
\begin{equation}
[G,G_c]\subseteq
G_c,\label{eq:semidirectconditionforbarG:pma-xuzhang200509}
\end{equation}
where $[G,G_c]=\{[A,B]\,|\,A\in G,\ B\in G_c\}$. Obviously, $G_c$
is an ideal Lie sub-algebra of $\bar G$. The subscript $c$ here
indicates a contribution to the construction of couplings. We also
require that the closure property between $G$ and $G_c$ under
matrix multiplication:
\begin{equation}
GG_c,G_cG\subseteq G_c,
\label{eq:conditiononLaxpairfrombarG:pma-xuzhang200509}\end{equation}
where $G_1G_2=\{AB\,|\, A\in G_1,\ B\in G_2\}$, to guarantee that
a Lax pair from the semi-direct sum $\bar G$ can generate a
coupling system. Note that the two different binary operations
were used in
 the above
closure properties in
(\ref{eq:semidirectconditionforbarG:pma-xuzhang200509}) and
(\ref{eq:conditiononLaxpairfrombarG:pma-xuzhang200509}).

Now choose a pair of new Lax matrices in the semi-direct sum $\bar
G$ of Lie algebras:
\begin{equation}
\bar U=U+U_c,\ \bar V=V+V_c,\ U_c,V_c\in
G_c,\label{eq:defofbarUandbarV:pma-xuzhang200509}
\end{equation}
and make a pair of enlarged discrete spatial matrix spectral
problems
\begin{equation}\left\{ \ba {l}
E\bar \phi =\bar U\bar \phi=\bar U(\bar u,\lambda )\bar \phi,
\vspace{2mm}
\\
\bar  \phi_{t}=\bar V\bar \phi =\bar V(\bar u,E\bar u,E^{-1}\bar
u\cdots ; \lambda )\bar \phi ,\ea \right.
\end{equation}
 where the matrix $U_c$ in $\bar U$ introduces additional
dependent variables and $\bar u$ consists of both the original
dependent variables and the additional dependent variables. In
addition, the matrix $U_c$ could depend on the spectral parameter
$\lambda $, and the matrix $V_c$ in $\bar V$ really does almost in
all cases. Based on the closure properties of $G$, $G_c$ and
between $G$ and $G_c$, it is easy to see that
\[ \ba {l}
\ (E\bar V)\bar U-\bar U\bar V =[(EV)U-UV ]\vspace{2mm}\\
\  +\{[(EV)U_c-U_cV]+[(EV_c)U-UV_c]+[(EV_c)U_c-U_cV_c]\}\in
G\inplus G_c.\ea
\]
Therefore, under $u_t=K(u)$, the corresponding enlarged discrete
zero curvature equation
\begin{equation} \bar U_t=(E\bar V)\bar U-\bar U\bar V\end{equation}
precisely presents
\begin{equation}
\left\{ \ba {l} U_t=(EV)U-UV,\vspace{2mm}
\\
U_{c,t}=[(EV)U_c-U_cV]+[(EV_c)U-UV_c]+[(EV_c)U_c-U_cV_c] . \ea
\right.\label{eq:infromsds:pma-xuzhang200509}
\end{equation}
The first equation above is equivalent to the lattice equation
(\ref{eq:ile:pma-xuzhang200509}), and hence,
 this is a coupling system for the lattice equation (\ref{eq:ile:pma-xuzhang200509}).

 The whole construction process above shows that
semi-direct sums of a given Lie algebra $G$ with new Lie algebras
provide a great choice of candidates of integrable couplings for
the lattice equation (\ref{eq:ile:pma-xuzhang200509}) generated
from the Lie algebra $G$.

\subsection{Realizations by particular semi-direct sums}

To shed light on the above general scheme of constructing coupling
systems, let us introduce the following particular class of
semi-direct sums of Lie algebras:
\begin{equation}\ba {l}
\bar G=G\inplus G_c,\
G=\Bigl\{\textrm{diag}(\,\underbrace{A,\cdots,A}_{\mu}\,,\,
\underbrace{0,\cdots,0}_{\nu -\mu +1}\,)
%\,\Bigl.\Bigr|\, A\in A_{r\times r}
\Bigr\} , \vspace{2mm}\\
 G_c=\left\{\, \left(
\begin{matrix}
0& & & & & \vspace{1mm}\\
 & \ddots  &    & & B_{ij} & \vspace{1mm}\\
 &  &0  & & & \vspace{1mm}\\
&  & & B_{\mu +1,\mu +1}&  & \vspace{1mm}\\
&  0& & & \ddots &  \vspace{1mm}\\
& & & & & B_{\nu +1,\nu +1}
\end{matrix}  \right)\, \right\},
\ea
 \label{eq:pcofsds:pma-xuzhang200509} \end{equation} where $A,B_{ii}$ are arbitrary
square matrices, $A$ is of the same order as $U$ and the
partitions of matrices in $G$ and $G_c$ are the same. Obviously,
$B_{ij},\  j\le \mu ,$ are square but $B_{ij},\ j\ge \mu +1,$ may
not; and all closure conditions of $G$, $G_c$ and between $G$ and
$G_c$ under matrix multiplication are satisfied.

Define \[ U_G:=\textrm{diag}(\,\underbrace{U,\cdots,U}_{\mu}
\,,\,\underbrace{0,\cdots,0}_{\nu -\mu +1}\,),\
V_G:=\textrm{diag}(\,\underbrace{V,\cdots,V}_\mu
\,,\,\underbrace{0,\cdots,0}_{\nu -\mu +1}\,).\]
 Note that $U_G$
and $V_G$
 in
$G$ generate the same lattice equation as $U$ and $V$, and thus
for integrable couplings, the corresponding enlarged spectral
matrices $\bar U$ and $\bar V$ in the semi-direct sum $ G\inplus
G_c $ can be chosen as
\begin{equation}\ba {l}
 \bar U=U_G+U_{G,c}:=
\left(
\begin{matrix}
U& & & & & \vspace{1mm}\\
 & \ddots  &    & & U_{ij} & \vspace{1mm}\\
 &  &U  & & & \vspace{1mm}\\
&  & & U_{\mu +1,\mu +1}&  & \vspace{1mm}\\
&  0& & & \ddots &  \vspace{1mm}\\
& & & & & U_{\nu +1,\nu +1}
\end{matrix}  \right)
\vspace{2mm}\\
 \bar V=V_G+V_{G,c}:=
 \left(
\begin{matrix}
V& & & & & \vspace{1mm}\\
 & \ddots  &    & & V_{ij} & \vspace{1mm}\\
 &  &V  & & & \vspace{1mm}\\
&  & & V_{\mu +1,\mu +1}&  & \vspace{1mm}\\
&  0& & & \ddots &  \vspace{1mm}\\
& & & & & V_{\nu +1,\nu +1}
\end{matrix}  \right)
 , \ea \nonumber
\end{equation}
where the first two matrices $U_G$ and $V_G$ play the $(U,V)$-part
and the second two matrices $U_{G,c}$ and $V_{G,c}$ play the
$(U_c,V_c)$-part in the pair of $\bar U$ and $\bar V$ defined in
(\ref{eq:defofbarUandbarV:pma-xuzhang200509}). It is not difficult
to see that the resulting coupling system
(\ref{eq:infromsds:pma-xuzhang200509}) becomes
\begin{equation}\left\{ \ba {l}
U_t=(EV)U-UV,\vspace{2mm}\\
 (U_{i+1,i+1})_t=(EV_{i+1,i+1})U_{i+1,i+1}-U_{i+1,i+1}V_{i+1,i+1}
 ,\ \mu \le i\le \nu
 ,\vspace{2mm}\\
U_{ij,t}=\D \sum_{k=i}^j [(EV_{ik})U_{kj}-U_{ik}V_{kj}],\ 1\le i<
j\le \nu +1,
 \ea \right.\label{eq:specificICfromsds:pma-xuzhang200509}
\end{equation}
where $U_{ii}=U$ and $ V_{ii}=V, \ 1\le i\le \mu$.

In particular, first, if we take
\[
 \bar U=
\left(
\begin{matrix}
U & U_{12}& \cdots & U_{1,\mu+1} \vspace{1mm}\\
 & \ddots      &  \ddots & \vdots \vspace{3mm}\\
 &  &U   & U_{\mu,\mu+1} \vspace{1mm}\\
0 &  & & 0
\end{matrix}  \right),\
\bar V= \left(
\begin{matrix}
V &V_{12} &\cdots & V_{1,\mu +1} \vspace{1mm}\\
 & \ddots      &  \ddots  & \vdots \vspace{3mm}\\
 &  &V   & V_{\mu,\mu +1}\vspace{1mm}\\
0 &  & & 0
\end{matrix}  \right),
 \]
then the coupling system
(\ref{eq:specificICfromsds:pma-xuzhang200509}) becomes
\begin{equation}
\left\{ \ba {l} U_t=(EV)U-UV,\vspace{2mm}\\
U_{ij,t}=\D \sum_{k=i}^j [(EV_{ik})U_{kj}-U_{ik}V_{kj}],\ 1\le i<
j\le \mu +1,
 \ea \right. \label{eq:1stspecificICfromsds:pma-xuzhang200509}
\end{equation}
where $U_{ii}=U,\ V_{ii}=V, \ 1\le i\le \mu,$ and
$U_{ii}=V_{ii}=0, \ i=\mu +1$.
 Second, if we take \[
 \bar U=
\left(
\begin{matrix}
U & U_{a_1}&\cdots  & U_{a_\nu } \vspace{1mm}\\
 & U      &   \ddots & \vdots\vspace{3mm}\\
 &  &  \ddots  & U_{a_1}\vspace{1mm}\\
 0 &  & & U
\end{matrix}  \right),\
 \bar V=
\left(
\begin{matrix}
V & V_{a_1}& \cdots & V_{a_\nu } \vspace{1mm}\\
 & V      & \ddots   & \vdots \vspace{3mm}\\
 &  &  \ddots  & V_{a_1}\vspace{1mm}\\
 0&  & & V
\end{matrix}  \right),
 \]
then the coupling system
(\ref{eq:specificICfromsds:pma-xuzhang200509}) becomes
\begin{equation}
\left\{\ba {l} U_t=(EV)U-UV,\vspace{2mm}\\
U_{a_i,t}=\D \sum_{k+l=i,k,l\ge 0}
[(EV_{a_k})U_{a_l}-U_{a_l}V_{a_k}],\ 1\le i\le \nu ,
 \ea \right.\label{eq:2ndspecificICfromsds:pma-xuzhang200509}
\end{equation}
where $U_{a_0}=U$ and $ V_{a_0}=V.$

We remark that here we have just presented one class of
semi-direct sums of Lie algebras, together with two specific
examples. It is interesting to construct other possible
realizations, especially those which could carry essential
information for keeping integrable properties of the original
lattice equations.

\subsection{Linearly dependent case on the spectral parameter}

 Let us now assume that the spatial spectral matrix
$U$ depends linearly on the spectral parameter $\lambda$ (see, for
example,
\cite{MaF-JMP1999,MaX-JPA2004,TomZ-CSF2005,YangXD-CSF2005}):
\begin{equation}
U=U(u,\lambda)=\lambda U_0+U_1,\  \frac {\partial U_0}{\partial
\lambda }=\frac {\partial U_1}{\partial \lambda }=0.
\end{equation}
%Note that not every $U$ belongs to this class (see, for example, \cite{Xu-CSF2003}).
Consider two specific examples of the enlarged spatial spectral
matrices introduced in the last sub-section: \begin{equation}
 \bar U_1=\left( \begin{matrix} U& U_a\vspace{1mm}\\
0 &0 \end{matrix} \right) ,\ \bar U_2=\left( \begin{matrix} U&
U_a\vspace{1mm}\\ 0 &U
\end{matrix} \right),\ \frac {\partial U_a}{\partial \lambda }=0 .
\end{equation}
Note that the sub-matrices $U_a$ in the above two enlarged spatial
spectral matrices could be of different sizes. As in the
continuous cases \cite{Ma-PLA2003,MaXZ-PLA2006}, suppose that
\[
\bar W_1= \left(\begin{matrix}
W&W_{a}\vspace{1mm}\\
0& 0
\end{matrix}\right),\
\bar W_2= \left(\begin{matrix}
W&W_{a}\vspace{1mm}\\
0& W
\end{matrix}\right)  \]
with
\begin{equation}
W=\sum_{i\ge 0}W_i\lambda ^{-i},\  W_{a}=\sum_{i\ge
-n_0}W_{a,i}\lambda ^{-i},\ \frac {\partial W_{i}}{\partial
\lambda }=0,\ \frac {\partial W_{a,i}}{\partial \lambda }=0,
\label{eq:formforWandW_a:pma-xuzhang200509}
\end{equation}
where $n_0\ge 0$ is a proper integer, solve the corresponding
enlarged discrete stationary zero curvature equations
\begin{equation}
 (E \bar W_i)\bar U_i-\bar U_i\bar W_i=0,\ i=1,2
,\label{eq:twospecificcasesofedszc:pma-xuzhang200509}\end{equation}
respectively.

 Then for each $m\ge 0$, choose
\[
 \bar V^{[m]}_1 = \left(\begin{matrix}
V^{[m]}&V_{a}^{[m]}\vspace{1mm}\\
0& 0
\end{matrix}\right)
= (\lambda ^m\bar W_1)_++\bar \Delta _m ,\ \bar \Delta_m=
\left(\begin{matrix}
\Delta _m&\Delta_{m,a}\vspace{1mm}\\
0& 0
\end{matrix}\right),\]
where $\Delta _m$ and $\Delta _{m,a}$ do not depend on $\lambda $
and satisfy \begin{equation} (E\Delta _m)U_0-U_0\Delta _m=0,\
U_0\Delta
_{m,a}=0;\label{eq:1stspecificconditionforDelta_mandDelta_{m,a}:pma-xuzhang200509}
\end{equation} and choose
\begin{equation} \bar V^{[m]}_2 = \left(\begin{matrix}
V^{[m]}&V_{a}^{[m]}\vspace{1mm}\\
0& V^{[m]}
\end{matrix}\right)
= (\lambda ^m\bar W_2)_++\bar \Delta _m ,\ \bar \Delta_m=
\left(\begin{matrix}
\Delta _m&\Delta_{m,a}\vspace{1mm}\\
0& \Delta_m
\end{matrix}\right),\nonumber
\end{equation}
where $\Delta _m$ and $\Delta _{m,a}$ do not depend on $\lambda $
and satisfy \begin{equation} (E\Delta _m)U_0-U_0\Delta _m=0,\
(E\Delta_{m,a})U_0-U_0\Delta
_{m,a}=0.\label{eq:2ndspecificconditionforDelta_mandDelta_{m,a}:pma-xuzhang200509}\end{equation}
The subscript $+$ above denotes to select the polynomial part in
$\lambda$. Based on
(\ref{eq:1stspecificICfromsds:pma-xuzhang200509}) and
(\ref{eq:2ndspecificICfromsds:pma-xuzhang200509}) and using
 (\ref{eq:twospecificcasesofedszc:pma-xuzhang200509}),
we can directly show
 that the enlarged discrete zero
curvature equations
\[ \bar U_{i,t_m}=(E\bar V^{[m]}_i)\bar U_i-\bar U_i\bar V^{[m]}_i,\ i=1,2,\]
namely,
\[
\left\{\ba {l} U_{t_m}=(EV^{[m]})U-UV^{[m]},
%U_{t_m}=(\Delta_{m})_x+[U_0,W_{m+1}]-[U_1,\Delta _{m}]=0,
\vspace{2mm}\\
U_{a,t_m}= (EV^{[m]})U_a-UV_a^{[m]}
,
 \ea \right.
 \ \textrm{and}\
\left\{\ba {l} U_{t_m}=(EV^{[m]})U-UV^{[m]},
%U_{t_m}=(\Delta_{m})_x+[U_0,W_{m+1}]-[U_1,\Delta _{m}]=0,
\vspace{2mm}\\
U_{a,t_m}= (EV^{[m]})U_a+(EV_a^{[m]})U
\vspace{2mm}\\
\qquad \quad \  -UV_a^{[m]}-U_aV^{[m]} ,
 \ea
\right.
\]
 present
\begin{equation}
\left\{\ba {l}
%U_{t_m}=(EV^{[m]})U-UV^{[m]},
U_{t_m}=(\Delta_{m})_x+[U_0,W_{m+1}]-[U_1,\Delta _{m}],
\vspace{2mm}\\
U_{a,t_m}=U_0W_{a,m+1}+(E\Delta _m)U_a-U_1\Delta_{m,a},
 \ea \right.\label{eq:1stmorespecificICfromsds:pma-xuzhang200509}
\end{equation}
and
\begin{equation}
\left\{\ba {l}
%U_{t_m}=(EV^{[m]})U-UV^{[m]},
U_{t_m}=(\Delta_{m})_x+[U_0,W_{m+1}]-[U_1,\Delta _{m}],
\vspace{2mm}\\
U_{a,t_m}=U_0W_{a,m+1}-(EW_{a,m+1})U_0\vspace{2mm}\\
\qquad \quad\  +(E\Delta _m)U_a-U_a\Delta _m+(E\Delta
_{m,a})U_1-U_1\Delta _{m,a},
 \ea \right.\label{eq:2ndmorespecificICfromsds:pma-xuzhang200509}
\end{equation}
respectively.

 We remark that these two enlarged hierarchies in
 (\ref{eq:1stmorespecificICfromsds:pma-xuzhang200509}) and
 (\ref{eq:2ndmorespecificICfromsds:pma-xuzhang200509}) share the
 enlarged discrete spectral problems
\[ E\bar \phi =\bar U _1\bar \phi,\  E\bar \phi =\bar U_2 \bar \phi, \]
respectively. Thus, all lattice equations in each of the two
enlarged hierarchies can possess infinitely many common conserved
densities except the original ones (see
\cite{TsuchidaUW-JMP1998,TsuchidaUW-JPA1999,ZhangC-CSF2002} for a
few concrete examples). Moreover, one can construct a specific
non-degenerate bilinear form on $\bar G$ with the invariance
property, to present Hamiltonian structures of the enlarged
lattice equations by a generalized trace identity. The detailed
analysis on those integrable properties will be left to a future
publication.

To sum up, each system of lattice equations in the hierarchy
(\ref{eq:1stmorespecificICfromsds:pma-xuzhang200509}) or
(\ref{eq:2ndmorespecificICfromsds:pma-xuzhang200509}) can provide
an integrable coupling for its first sub-system of lattice
equations. In the next section, we will only discuss two examples
of constructing enlarged lattice hierarchies, in the generalized
Toda case presented in \cite{MaX-JPA2004}.

\section{Integrable couplings of the generalized Toda hierarchy}

\subsection{The generalized Toda equations}
Let us here recall the generalized Toda hierarchy
\cite{MaX-JPA2004}. The corresponding discrete spatial spectral
problem reads \be E\phi =U(u,\lambda )\phi,\ U(u,\lambda
)=\left(\ba {cc}0& 1 \vspace{2mm}\\  (\alpha \la +\beta )r & \la
+s \ea \right),\ u=\left(\ba {c}r \vspace{2mm}\\  s  \ea \right),
 \label{eq:spofgToda:pma-xuzhang200509}\ee
where $\la $ is a spectral parameter, and $\al $ and $\beta $ are
two arbitrary constants satisfying $\alpha ^2+\beta ^2\ne 0$. When
$\alpha =0$ and $\beta =-1$,
(\ref{eq:spofgToda:pma-xuzhang200509}) becomes the Toda spectral
problem \cite{Tu-JPA1990}.

Its stationary discrete zero curvature equation \be (EW )U-UW =0
%\label{eq:dszce:pma-xuzhang200509}
\ee has the solution \be W
=\left(\ba {cc}a& b \vspace{2mm}\\  (\alpha \la +\beta ) c&-a \ea
\right), \label{eq:solutiontoszce:pma-xuzhang200509} \ee with
 \[
a=\sum_{i\ge 0} a_i\la ^{-i},\ b=\sum_{i\ge 0} b_i\la ^{-i},\
c=\sum_{i\ge 0 } c_i\la ^{-i},
 \] where the coefficients are defined by the initial conditions:
 \[ a_0=-\frac 12,\ b_0=0,\ c_0=0,
 %\label{eq:relationfora0b0c0:pma-xuzhang200509}
 \]
and the recursion relation: \begin{equation}
\left\{ \ba {l} c_{i+1}-rb^{(1)}_{i+1}=0, \vspace{2mm}\\
 b^{(1)}_{i+1}+sb^{(1)}_i+(a^{(1)}_i+a_i) =0,
\vspace{2mm}\\
(a^{(1)}_{i+1}-a_{i+1})+s (a^{(1)}_i-a _i) \vspace{2mm}\\
\ +\al (rb_{i+1}-c^{(1)}_{i+1})+\beta (rb_{i}-c^{(1)}_{i})=0,
 \ea \right.\ i\ge 0,
\label{eq:rlfora_mb_mc_m:pma-xuzhang200509} \end{equation} which
are all difference polynomials in $u$ with respect to the lattice
variable $n$.
 Under the initial-value conditions
\[ a_{1}|_{u=0}=c_1|_{u=0}=0, \
a_i|_{u=0}=b_i|_{u=0}=c_i|_{u=0}=0,\ i\ge 2, \] the recursion
relation (\ref{eq:rlfora_mb_mc_m:pma-xuzhang200509}) uniquely
determine the lattice functions $a_i,b_i$ and $c_i$, $i\ge 1$.
 The first few
lattice functions are  \[
 \left\{
 \ba {l} a_1=\al r,\ b_1=1,\ c_1=r; \vspace{2mm}\\
a_2=-\al ^2r^{(1)}r-\alpha ^2r^2-\al ^2rr^{(-1)} -\alpha rs -\al
rs^{(-1)}+\beta r,
\vspace{2mm}\\
b_2=-\al r-\al r^{(-1)} -s^{(-1)},\ c_2=-rs -\al r^2-\al rr^{(1)}.
 \ea \right.
%\label{eq:firstfewformulafora_mb_mc_m:pma-xuzhang200509}
 \]

As usual, choose that \bea
V_m&=& \left(\ba {cc}(\la ^ma)_+& (\la ^mb)_+ \vspace{2mm}\\
(\alpha \la +\beta )(\la ^mc)_+ & -(\la ^ma)_+ \ea \right),\ m\ge
0. \eea Then it follows from
(\ref{eq:rlfora_mb_mc_m:pma-xuzhang200509}) that
\[ (EV_m)U-UV_m=
\left(\ba {cc} 0 & -b_{m+1}^{(1)} \vspace{2mm}\\
(\alpha \la +\beta )c_{m+1} & \beta
(c_m^{(1)}-rb_m)-s(a_m^{(1)}-a_m) \ea \right).
\]
Take a
modification \[ \Delta _m=\left(\ba {cc} b_{m+1} & 0\vspace{2mm}\\
0&0 \ea \right), \]  and define the temporal spectral matrices \be
V^{[m]}=V_m+\Delta _m,\ m\ge 0 .
 \label{eq:defofV^{[m]}:pma-xuzhang200509} \ee Then, a direct
calculation leads to the following matrix:
\[ (EV^{[m]})U-UV^{[m]}= \left(\ba {cc} 0 &  0 \vspace{2mm}\\
(\alpha \la +\beta )(c_{m+1} -rb_{m+1})&\beta
(c_m^{(1)}-rb_m)-s(a_m^{(1)}-a_m) \ea \right). \]
 This is
consistent with $U_{t_m}$, and thus, making the evolution laws
 \be
\phi_{t_m}=V^{[m]}\phi,\ m\ge 0,
\label{eq:timesp:pma-xuzhang200509} \ee
 the compatibility
conditions \[ U_{t_m}=(EV^{[m]})U-UV^{[m]},\ m\ge 0, \] of the
discrete spatial spectral problem
(\ref{eq:spofgToda:pma-xuzhang200509}) and the associated discrete
temporal spectral problems (\ref{eq:timesp:pma-xuzhang200509})
give rise to the following hierarchy of lattice equations \be
\left\{ \ba {l} r_{t_m}=c_{m+1}-rb_{m+1},
\vspace{2mm}\\
s_{t_m}=-\al (c_{m+1}^{(1)}-rb_{m+1})+(a_{m+1}^{(1)}-a_{m+1}) ,
\ea \right. \ m\ge 0. \label{eq:gTodah:pma-xuzhang200509}\ee This
generalized Toda hierarchy is Liouville integrable
\cite{MaX-JPA2004}, and its Hamiltonian structure leads to
infinitely many conservation laws and symmetries for every system
in the hierarchy.

Obviously, the first nonlinear lattice equation in the hierarchy
%(\ref{eq:gTodah:pma-xuzhang200509})
 is \be \left\{\ba {l}
r_{t_1}=r(s^{(-1)}-s)+\al r(r^{(-1)}-r^{(1)}) ,\vspace{2mm}\\
s_{t_1}=\al s(r-r^{(1)})+\beta (r^{(1)}-r). \ea \right.
\label{eq:firstlatticeindsh:pma-xuzhang200509} \ee
 When $\al =0$
and $\beta =-1$, (\ref{eq:firstlatticeindsh:pma-xuzhang200509})
becomes the Toda lattice equation \cite{Toda-book1989}: \be
r_{t_1}=r(s^{(-1)}-s) ,\ s_{t_1}=r-r^{(1)};
\label{eq:Toda:pma-xuzhang200509} \ee
 and when $\al =1$ and
$\beta =0$, (\ref{eq:firstlatticeindsh:pma-xuzhang200509}) becomes
the following lattice equation presented in \cite{MaX-IJTP2004}:
\be r_{t_1}=r(s^{(-1)}-s)+ r(r^{(-1)}-r^{(1)}) ,\
s_{t_1}=s(r-r^{(1)}). \label{eq:2ndsublattice:pma-xuzhang200509}
\ee  The lattice equation
(\ref{eq:2ndsublattice:pma-xuzhang200509}) is linearly independent
of the Toda lattice equation (\ref{eq:Toda:pma-xuzhang200509}).
There exist a voluminous literature on the Toda lattice equation,
and its generalizations and solution structures (for example, see
\cite{Manakov-SPJETP1975}-\cite{MaM-PA2004}).

\subsection{Integrable couplings from specific semi-direct sums}

The generalized Toda spectral problem
(\ref{eq:spofgToda:pma-xuzhang200509}) linearly depends on the
spectral parameter $\lambda $, and thus we can write
\begin{equation}U=\left(\ba
{cc}0& 1 \vspace{2mm}\\  (\alpha \la +\beta )r & \la +s \ea
\right)
=
U_0\lambda +U_1,\ U_0=\left( \begin{matrix}0&0 \vspace{2mm}
\\ \alpha r &1\end{matrix}\right), \ U_1=\left(
\begin{matrix}0&1 \vspace{2mm} \\ \beta r&s \end{matrix}\right).
 \end{equation}
We will also see that there is a difference between the two cases
of $\alpha =0$ and $\alpha \ne 0$ in computing integrable
couplings.

Let us first consider the semi-direct sum of Lie algebras of
$3\times 3$ matrices:
 \begin{equation} G\inplus G_c,\ G=\left\{ \left.\left( \begin{matrix}A &0
\vspace{2mm}
\\0&0 \end{matrix}\right) \right| A\in \mathbb{ C}[\lambda , \lambda ^{-1}]\otimes M_{2\times 2}\right\} ,\
G_c=\left\{ \left. \left( \begin{matrix}0 &B \vspace{2mm}
\\0&0 \end{matrix}\right) \right| B\in \mathbb{ C}[\lambda , \lambda ^{-1}]\otimes M_{2\times
1}\right\},
\nonumber
 \end{equation}
 where $\mathbb{ C}[\lambda , \lambda ^{-1}]\otimes M_{m\times n}=\textrm{span}\{\,\lambda ^kA\,|\, k\in \mathbb{ Z},\ A\in M_{m\times
 n}\,\}$.
 In this case, $G_c$ is an Abelian ideal of
$ G\inplus G_c$. We define the corresponding enlarged spatial
spectral matrix as
\begin{equation}
\bar U= \bar U(\bar u,\lambda )=\left( \begin{matrix}U &U_a
\vspace{2mm}
\\0&0 \end{matrix}\right)\in G\inplus G_c ,\ U_a=U_a(v)=\left( \begin{matrix}v_1  \vspace{2mm}
\\v_2 \end{matrix}\right),
\end{equation}
where $v_1$ and $v_2$ are new dependent variables and
\begin{equation} v=(v_1,v_2)^T,\ \bar u=(u^T,v^T)^T= (r,s,v_1,v_2)^T.
\nonumber
%\label{eq:defof1stvandbaru:pma-xuzhang200509}
\end{equation}

Upon setting
\begin{equation}
\bar W= \left( \begin{matrix}
W &W_a \vspace{2mm}
\\0&0 \end{matrix}\right),\
W_a=W_a(\bar u,\lambda)=\left( \begin{matrix} e  \vspace{2mm}
\\ f \end{matrix}\right),\nonumber
\end{equation}
where $W$ is a solution to $(EW)U-UW=0$, defined by
(\ref{eq:solutiontoszce:pma-xuzhang200509}), the corresponding
enlarged discrete stationary zero curvature equation $ (E\bar
W)\bar U-\bar U\bar W=0$ becomes \begin{equation}
 (EW)U_a-UW_a=0, \label{eq:dzcefor1stcase:pma-xuzhang200509}
\end{equation} which is equivalent to
\[W_a=WU^{-1}U_a,\]
namely,
\begin{equation}\left\{\ba {l}
f=a^{(1)}v_1+b^{(1)}v_2,\vspace{2mm}\\
(\alpha \lambda +\beta )re= (\alpha \lambda +\beta
)c^{(1)}v_1-a^{(1)}v_2-(\lambda +s) f.
 \ea \right.\label{eq:relationforefgin1stcase:pma-xuzhang200509}
\end{equation}
This system determines a solution for $e$ and $f$ as follows
\[ e=\sum_{i\ge -n_0}e_i\lambda ^{-i},\ f=\sum_{i\ge 0}f_i\lambda ^i, \]
where $n_0=1$ if $\alpha =0$ and $n_0=0$ if $\alpha \ne 0$ (see
(\ref{eq:formforWandW_a:pma-xuzhang200509}) for introduction of
$n_0$).
% Expand $W_a$ as \begin{equation} W_a=\sum_{i\ge -n_0}W_{a,i}\lambda ^{-i}  \end{equation}
Now define the enlarged temporal spectral matrix as
\begin{equation} \bar V^{[m]}= \left(
\begin{matrix}V^{[m]} &V_a^{[m]} \vspace{2mm}
\\0&0 \end{matrix}\right),\
V_a^{[m]}=(\lambda ^mW_a)_+ +\Delta _{m,a},\ m\ge
0,\label{eq:1stchoiceforbarV^{[m]}:pma-xuzhang200509}\end{equation}
where $V^{[m]}$ is defined as in
(\ref{eq:defofV^{[m]}:pma-xuzhang200509}). To satisfy
(\ref{eq:1stspecificconditionforDelta_mandDelta_{m,a}:pma-xuzhang200509}),
choose $\Delta _{m,a}$ as \begin{equation}
 \Delta _{m,a}=
 \left( \begin{matrix}h_{m}
\vspace{2mm}
\\ -\alpha rh_{m}\end{matrix}\right),\ h_m\  -\  \textrm{arbitrary},\ m\ge 0.
\label{eq:choiceforDelta_mfor1stcase:pma-xuzhang200509}
\end{equation}
 Then based on
(\ref{eq:relationforefgin1stcase:pma-xuzhang200509}), we can
compute that
\[ \ba {l}
(EV^{[m]})U_a-UV_a^{[m]}= (EV_m)U_a-U(\lambda ^mW_a)_+ +(E\Delta
_m)U_a-U_1\Delta_{m,a}\vspace{2mm}\\
= \left( \begin{matrix}0  \vspace{2mm}
\\ -\alpha c_{m+1}^{(1)}v_1+\alpha r e_{m+1}+f_{m+1} \end{matrix}\right)
+\left( \begin{matrix} b_{m+1}^{(1)}v_1 \vspace{2mm}
\\ 0 \end{matrix}\right)
-\left( \begin{matrix} -\alpha r h_m  \vspace{2mm}
\\ \beta r h_{m}-\alpha rsh_m  \end{matrix}\right) \vspace{2mm}\\
= \left( \begin{matrix} b_{m+1}^{(1)}v_1 + \alpha r h_m
\vspace{2mm}
\\ -\alpha c_{m+1}^{(1)}v_1+\alpha r e_{m+1}+f_{m+1} - \beta r h_{m}+\alpha rsh_m
\end{matrix}\right),\ m\ge 0.
 \ea  \]
Therefore, the $m$-th enlarged discrete zero curvature equation
\[\bar U_{t_m}=(E \bar V^{[m]})\bar U-\bar U\bar V^{[m]}\]
leads to
\begin{equation}
v_{t_m}= \left(\begin{matrix} v_1 \vspace{2mm}
\\ v_2 \end{matrix}
\right)_{t_m}=S_m(u,v) = \left(\begin{matrix} b_{m+1}^{(1)}v_1 +
\alpha r h_m \vspace{2mm}
\\
-\alpha c_{m+1}^{(1)}v_1+\alpha r e_{m+1}+f_{m+1} - \beta r
h_{m}+\alpha rsh_m
\end{matrix} \right),
\label{eq:1stVFforgToda:pma-xuzhang200509}\end{equation}
 together
with the $m$-th generalized Toda equation in
(\ref{eq:gTodah:pma-xuzhang200509}). Therefore, we obtain a
hierarchy of coupling systems defined by
(\ref{eq:1stmorespecificICfromsds:pma-xuzhang200509}):
\begin{equation}
\left. \ba {l} \bar u_{t_m}=\left( \ba {l} u\vspace{2mm}\\
v
 \ea \right)_{t_m}=
 \bar K_m(u)
 =
 \left( \ba {c} K_m(u)\vspace{2mm}\\
S_{m}(u,v)
 \ea \right),\ea \right. \  m\ge 0 \label{eq:1stICforgToda:pma-xuzhang200509}\end{equation}
for the generalized Toda hierarchy
(\ref{eq:gTodah:pma-xuzhang200509}).

Let us second consider the semi-direct sum of Lie algebras of
$4\times 4$ matrices:
 \[ G\inplus G_c,\ G=\left\{ \left.\left( \begin{matrix}A &0
\vspace{2mm}
\\0&A \end{matrix}\right) \right| A\in \mathbb{ C}[\lambda , \lambda ^{-1}]\otimes M_{2\times 2}\right\} ,\
G_c=\left\{ \left. \left( \begin{matrix}0 &B \vspace{2mm}
\\0&0 \end{matrix}\right) \right| B\in \mathbb{ C}[\lambda , \lambda ^{-1}]\otimes M_{2\times 2}\right\}.
 \]
 In this case, $G_c$ is an Abelian ideal of $G\inplus G_c$, too.
We define the corresponding enlarged spatial spectral matrix as
\begin{equation}
\bar U= \bar U(\bar u,\lambda )=\left( \begin{matrix}U &U_a
\vspace{2mm}
\\0&U \end{matrix}\right)\in G\inplus G_c ,\ U_a=U_a(v)=\left( \begin{matrix}v_1 &v_2  \vspace{2mm}
\\v_3&v_4 \end{matrix}\right),
\end{equation}
where $v_i$, $1\le i\le 4$, are new dependent variables and
\begin{equation}
v=(v_1,v_2,v_3,v_4)^T,\ \bar u=(u^T,v^T)^T=
(r,s,v_1,v_2,v_3,v_4)^T. \nonumber
%\label{eq:defof2ndvandbaru:pma-xuzhang200509}
\end{equation}

If we set
\[
 \bar W= \left( \begin{matrix}W &W_a \vspace{2mm}
\\0&W \end{matrix}\right),\ W_a=W_a(\bar u,\lambda)=\left( \begin{matrix}e &f \vspace{2mm}
\\ g&-e \end{matrix}\right),
\]
where $W$ is a solution to $(EW)U-UW=0$, defined by
(\ref{eq:solutiontoszce:pma-xuzhang200509}), then the
corresponding enlarged discrete stationary zero curvature equation
$ (E\bar W)\bar U-\bar U\bar W=0$ becomes \begin{equation}
(EW)U_a+(EW_a)U-UW_a-U_aW=0,
\label{eq:dzcefor2ndcase:pma-xuzhang200509}
\end{equation}
 which is equivalent to
\begin{equation}\left\{\ba {l}
(e^{(1)}+e)+(\lambda +s)f^{(1)} +(a^{(1)}+a)v_2+b^{(1)}v_4-bv_1=0,\vspace{2mm}\\
-(\alpha \lambda +\beta )r(e^{(1)}+e) - (\lambda +s)g+ (\alpha
\lambda +\beta )(c^{(1)}v_1-cv_4)-(a^{(1)}+a)v_3
=0,\vspace{2mm}\\
g^{(1)}-(\lambda +s)(e^{(1)}-e)-(\alpha \lambda +\beta
)(rf-c^{(1)}v_2) -(a^{(1)}-a)v_4-bv_3=0.
 \ea \right. \label{eq:relationforefgin2ndcase:pma-xuzhang200509}
\end{equation}
This system can determine a solution for $e$, $f$ and $g$ as
follows
\[ e=\sum_{i\ge 0}e_i\lambda ^{-i},\ f=\sum_{i\ge 0}f_i\lambda ^{-i},\  g=\sum_{i\ge 0}g_i\lambda ^{-i}. \]
Now, we define the enlarged temporal spectral matrix as
\begin{equation} \bar V^{[m]}= \left(
\begin{matrix}V^{[m]} &V_a^{[m]} \vspace{2mm}
\\0&V^{[m]} \end{matrix}\right),\
V_a^{[m]}=(\lambda ^mW_a)_+ +\Delta _{m,a},\ m\ge
0,\label{eq:2ndchoiceforbarV^{[m]}:pma-xuzhang200509}\end{equation}
where $V^{[m]}$ is defined as in
(\ref{eq:defofV^{[m]}:pma-xuzhang200509}). To satisfy
(\ref{eq:2ndspecificconditionforDelta_mandDelta_{m,a}:pma-xuzhang200509}),
choose $\Delta _{m,a}$ as \begin{equation}
 \Delta _{m,a}=
 \left( \begin{matrix} h_m &0
\vspace{2mm}
\\ -\alpha r h_m &0 \end{matrix}\right),\ h_m\  -\  \textrm{arbitrary},\ m\ge 0.
\label{eq:choiceforDelta_mfor2ndcase:pma-xuzhang200509}
\end{equation}
 Then based on (\ref{eq:relationforefgin2ndcase:pma-xuzhang200509}), we
can compute that
\[\ba {l}
(EV^{[m]})U_a+(EV_a^{[m]})U-UV_a^{[m]}-U_aV^{[m]}\vspace{2mm}\\
=[(EV_m)U_a+(E(\lambda ^mW_a)_+)U-U(\lambda
^mW_a)_+-U_aV_m]\vspace{2mm}\\
\quad +[(E\Delta _m)U_a-U_a\Delta _m]+[(E\Delta _{m,a})U_1-U_1\Delta _{m,a}]\vspace{2mm}\\
= \left(
\begin{matrix}-\alpha c_{m+1}v_2+\alpha r f^{(1)}_{m+1} &f_{m+1}^{(1)} \vspace{2mm}
\\ \alpha c_{m+1}^{(1)}v_1 -\alpha c_{m+1}v_4-\alpha r (e_{m+1}^{(1)}+e_{m+1})-g_{m+1}&
\alpha c_{m+1}^{(1)}v_2-\alpha r f_{m+1}-(e_{m+1}^{(1)}-e_{m+1}
 \end{matrix}\right)
\vspace{2mm}\\
\quad + \left(
\begin{matrix}(b_{m+1}^{(1)}-b_{m+1})v_1 &b_{m+1}^{(1)}v_2 \vspace{2mm}
\\ -b_{m+1}v_3&0 \end{matrix}\right)
+ \left(
\begin{matrix}\alpha rh_m&h_m^{(1)} \vspace{2mm}
\\ r(\alpha s-\beta) h_m& -\alpha r^{(1)}h_m^{(1)}
\end{matrix}\right)\vspace{2mm}\\
= \left(
\begin{matrix}
-\alpha c_{m+1}v_2+\alpha r
f^{(1)}_{m+1}+(b_{m+1}^{(1)}-b_{m+1})v_1+\alpha rh_m,
\vspace{2mm}\\
\alpha c_{m+1}^{(1)}v_1 -\alpha c_{m+1}v_4-\alpha r
(e_{m+1}^{(1)}+e_{m+1})-g_{m+1} -b_{m+1}v_3 + r(\alpha s-\beta)
h_m,
\end{matrix}\right. \vspace{2mm}\\
\quad\  \left.\begin{matrix} f_{m+1}^{(1)}+b_{m+1}^{(1)}v_2
+h_m^{(1)}
  \vspace{2mm}\\
 \alpha c_{m+1}^{(1)}v_2-\alpha r f_{m+1}-(e_{m+1}^{(1)}-e_{m+1})
 -\alpha
r^{(1)}h_m^{(1)}
\end{matrix}\right)
 .
 \ea  \]
Then the $m$-th enlarged discrete zero curvature equation
\[\bar U_{t_m}=(E \bar V^{[m]})\bar U-\bar U\bar V^{[m]}\]
leads to
\begin{equation}
\ba{l}
 v_{t_m}= (v_1 ,v_2,v_3,v_4)^T
_{t_m}=
T_m(u,v) \vspace{2mm}\\
 = \left(\begin{matrix} -\alpha c_{m+1}v_2+\alpha r
f^{(1)}_{m+1}+(b_{m+1}^{(1)}-b_{m+1})v_1+\alpha rh_m
 \vspace{2mm}
\\ f_{m+1}^{(1)}+b_{m+1}^{(1)}v_2 +h_m^{(1)}
 \vspace{2mm}
\\
\alpha c_{m+1}^{(1)}v_1 -\alpha c_{m+1}v_4-\alpha r
(e_{m+1}^{(1)}+e_{m+1})-g_{m+1} -b_{m+1}v_3 + r(\alpha s-\beta)
h_m
\vspace{2mm}\\
\alpha c_{m+1}^{(1)}v_2-\alpha r f_{m+1}-(e_{m+1}^{(1)}-e_{m+1})
 -\alpha
r^{(1)}h_m^{(1)}
\end{matrix}\right),
 \ea \label{eq:2ndVFforgToda:pma-xuzhang200509}
\end{equation} together with the $m$-th generalized
Toda equation in (\ref{eq:gTodah:pma-xuzhang200509}). Therefore,
we obtain a hierarchy of coupling systems defined by
(\ref{eq:2ndmorespecificICfromsds:pma-xuzhang200509}):
\begin{equation}
\left. \ba {l} \bar u_{t_m}=\left( \ba {l} u\vspace{2mm}\\
v
 \ea \right)_{t_m}=
 \bar K_m(u)
 =
 \left( \ba {c} K_m(u)\vspace{2mm}\\
T_{m}(u,v)
 \ea \right),\ea \right. \  m\ge 0 \label{eq:2ndICforgToda:pma-xuzhang200509}\end{equation}
for the generalized Toda  hierarchy
(\ref{eq:gTodah:pma-xuzhang200509}).

\subsection{Illustrative examples}

We now work out two concrete examples as follows, one in each of
the two above cases.

 {\bf Case of $\alpha =1$ and $\beta=0$:}
  Let us first compute an
example of the hierarchy
(\ref{eq:1stICforgToda:pma-xuzhang200509}). Assume that $\alpha
=1$ and $\beta=0$ for convenience, which corresponds to the
lattice hierarchy presented in \cite{MaX-IJTP2004}. In this case,
we have $n_0=0$ in (\ref{eq:formforWandW_a:pma-xuzhang200509}).

It directly follows from
(\ref{eq:relationforefgin1stcase:pma-xuzhang200509}) that
\[\left\{ \ba {l}
f_{i}=a_i^{(1)}v_1+b_i^{(1)}v_2,
\vspace{2mm}\\
re_{i+1}=c_{i+1}^{(1)}v_1-a_{i}^{(1)}v_2-f_{i+1}-sf_{i},   \ea
\right.
 \]
 where $i\ge 0$ and $ re_0=c_0^{(1)}v_1-f_0.$
We can then obtain that
\[
\left\{\ba {l} f_0=-\frac 12 v_1,\
f_1=r^{(1)}v_1+v_2,\vspace{2mm}\\ f_2=
-[r^{(2)}r^{(1)}+(r^{(1)})^2+
r^{(1)}r+r^{(1)}s^{(1)}+r^{(1)}s ]v_1-(r^{(1)}+r+s)v_2;\vspace{2mm}\\
re_0=\frac 12 v_1,\ re_1= \frac 12 sv_1-\frac 12 v_2,\
re_2=r^{(1)}rv_1+rv_2.
 \ea \right.
\]

If we choose
\[h_1= 2\xi f_0f_1=-\xi v_1(r^{(1)}v_1+v_2),\ \xi=\textrm{const.}, \]
 then the vector-valued function $S_1$ defined by
(\ref{eq:1stVFforgToda:pma-xuzhang200509}) becomes
\[ S_1(u,v)=
\left(\begin{matrix} -(r^{(1)}+r+s)v_1-\xi rv_1(r^{(1)}v_1+v_2)
 \vspace{2mm}
\\ -r^{(1)}sv_1-(r^{(1)}+s)v_2-\xi rsv_1(r^{(1)}v_1+v_2)
\end{matrix}\right).
\]
Therefore, the integrable coupling of the generalized Toda lattice
equation (\ref{eq:2ndsublattice:pma-xuzhang200509}), defined by
(\ref{eq:1stICforgToda:pma-xuzhang200509}), reads as
\begin{equation}
\left\{\ba {l} r_{t_1}=r(s^{(-1)}-s)+r(r^{(-1)}-r^{(1)}), \vspace{2mm}\\
s_{t_1}=s(r-r^{(1)}), \vspace{2mm}\\
v_{1,t_1}= -(r^{(1)}+r+s)v_1-\xi rv_1(r^{(1)}v_1+v_2),
 \vspace{2mm}
\\ v_{2,t_1}=
-r^{(1)}sv_1-(r^{(1)}+s)v_2-\xi rsv_1(r^{(1)}v_1+v_2),
 \ea \right.
\end{equation}
the second sub-system of which is nonlinear with respect to both
sub-sets of dependent variables when $\xi \ne 0$.

{\bf Case of $\alpha=0$ and $\beta =1$:}
 Let us second compute an
example of the hierarchy
(\ref{eq:2ndICforgToda:pma-xuzhang200509}). Assume that $\alpha
=0$ and $\beta=1$ for convenience, which corresponds to the Toda
lattice hierarchy. In this case, we have $n_0=1$ in
(\ref{eq:formforWandW_a:pma-xuzhang200509}).

We take the initial set of functions as follows:
\[f_0=g_0=0,\ e_0=-\frac 1 2.\]
Obviously from
(\ref{eq:relationforefgin2ndcase:pma-xuzhang200509}), we can have
\[\left\{ \ba {l}
f^{(1)}_{i+1}=-sf_i^{(1)}-(e_i^{(1)}+e_i)-(a_i^{(1)}+a_i)v_2-b_i^{(1)}v_4+b_iv_1,
\vspace{2mm}\\
g_{i+1}=-sg_i-r(e_i^{(1)}+e_i)-(a_i^{(1)}+a_i)v_3
+c_i^{(1)}v_1-c_iv_4, \vspace{2mm}\\
e_{i+1}^{(1)}-e_{i+1}=-s(e_i^{(1)}-e_i)+g_i^{(1)}-rf_i-(a_i^{(1)}-a_i)v_4
+c_i^{(1)}v_2-b_iv_3,
  \ea \right.
 \]
 where $i\ge 0.$
It then follows that
\[
\left\{\ba {l} f_1=1+v_2^{(-1)}, \ g_1=r+v_3,\
e_1=v_3+r(1+v_2^{(-1)});\vspace{2mm}\\
f_2=-(s^{(-1)}+r)(1+v_2^{(-1)})-v_3-v_3^{(-1)}-r^{(-1)}(1+v_2^{(-2)})-v_4^{(-1)}+v_1^{(-1)},\vspace{2mm}\\
g_2=-s(r+v_3)
-r[v_3^{(1)}+v_3+r^{(1)}(1+v_2)+r(1+v_2^{(-1)})]+r^{(1)}v_1-rv_4,\vspace{2mm}\\
e_2^{(1)}-e_2=(1-
s)(r^{(1)}-r+v_3^{(1)}-v_3+r^{(1)}v_2-rv_2^{(-1)}) .
 \ea \right.
\]
We can use the inverse formula
(\ref{eq:ivofe-1:pma-xuzhang200509}) to compute $e_2$ here, but as
we will see, this is not necessary for computing the corresponding
integrable coupling.

Now if we choose
\[ h_1=\eta v_1e_1= \eta v_1[v_3+r(1+v_2^{(-1)})],\ \eta=\textrm{const.},\]
 then the vector-valued function $T_1$
defined by (\ref{eq:2ndVFforgToda:pma-xuzhang200509}) becomes
\[ T_1(u,v)=
\left(\begin{matrix} -(s-s^{(-1)})v_1
 \vspace{3mm}
\\-sv_2 -(s+r^{(1)})(1+v_2)-r(1+v_2^{(-1)})+v_1-v_3^{(1)}
 \vspace{1mm}
\\
\qquad \qquad \ -v_3-v_4 +\eta v_1^{(1)}[v_3^{(1)}+r^{(1)}(1+v_2)]
\vspace{3mm}\\
(s^{(-1)}+s)v_3+rs+r[v_3^{(1)}+v_3+r^{(1)}(1+v_2)+r(1+v_2^{(-1)})]
\vspace{1mm}\\
\qquad \quad\ \  \qquad  -r^{(1)}v_1+rv_4 -\eta
rv_1[v_3+r(1+v_2^{(-1)})]
\vspace{3mm}\\
(s-1)(r^{(1)}-r+v_3^{(1)}-v_3+r^{(1)}v_2-rv_2^{(-1)})
\end{matrix}\right).
\]
Therefore, the integrable coupling of the Toda lattice equation
(\ref{eq:Toda:pma-xuzhang200509}), defined by
(\ref{eq:2ndICforgToda:pma-xuzhang200509}), reads as
\begin{equation}
\left\{\ba {l} r_{t_1}=r(s^{(-1)}-s), \ s_{t_1}=r^{(1)}-r,
\vspace{3mm}
\\
v_{1,t_1}=-(s-s^{(-1)})v_1,
 \vspace{3mm}
\\ v_{2,t_1}=-sv_2-(s+r^{(1)})(1+v_2)-r(1+v_2^{(-1)})+v_1-v_3^{(1)}
 \vspace{1mm}
\\
\qquad \quad -v_3-v_4 + \eta v_1^{(1)}[v_3^{(1)}+r^{(1)}(1+v_2)],
 \vspace{3mm}
\\
v_{3,t_1}=(s^{(-1)}+s)v_3+rs+r[v_3^{(1)}+v_3+r^{(1)}(1+v_2)+r(1+v_2^{(-1)})]
\vspace{1mm}\\
\qquad \quad -r^{(1)}v_1+rv_4 -\eta rv_1[v_3+r(1+v_2^{(-1)})],
 \vspace{3mm}
\\
v_{4,t_1}=(s-1)(r^{(1)}-r+v_3^{(1)}-v_3+r^{(1)}v_2-rv_2^{(-1)}),
\ea \right.
\end{equation}
the second sub-system of which is nonlinear with respect to both
sub-sets of dependent variables when $\eta \ne 0$.

\section{Conclusions and remarks}

A feasible approach to construct integrable couplings of discrete
soliton equations has been proposed by taking advantage of
semi-direct sums of Lie algebras, and the resulting theory has
been applied to the generalized Toda hierarchy of lattice
equations to generate integrable couplings for the hierarchy. The
key point in our generating scheme is to establish a relation
between semi-direct sums of Lie algebras and integrable couplings
of discrete soliton equations. The underlying discrete matrix
spectral problems are generated from semi-direct sums of Lie
algebras, and the discrete Lax spectral matrices associated with
given soliton equations play the non-ideal part in the semi-direct
sums.

In our analysis of the two specific semi-direct sums, we have seen
that there is always an arbitrary modified term $\Delta _{n,a}$.
This indicates that higher order matrix spectral problems have
more degrees of freedom in generating integrable systems. On the
other hand, in all additional spectral sub-matrices such as
$U_{a_i}$, one can take their dependence on the spectral parameter
into consideration, and this will bring much more diverse
integrable couplings.

There are also other questions about integrable properties of the
resulting enlarged lattice equations, even in the case where
additional spectral sub-matrices are independent of the spectral
parameter. For example, can we solve the enlarged lattice
equations by the inverse scattering transform? The class of Lie
algebras in (\ref{eq:pcofsds:pma-xuzhang200509}) provides a few
realizations of semi-direct sums of Lie algebras. Other possible
realizations are still interesting, especially those which could
carry significant information about integrable properties.
Reductions of the presented cases of semi-direct sums, which keep
the uniqueness property of discrete spectral problems (see
\cite{MaF-JMP1999}), could be good examples.

We would especially like to emphasize that we have been
considering the problem of integrable couplings and the key is
semi-direct sums of Lie algebras. The initial Lie algebras $G$
associated with given integrable systems in our construction can
be simple (e.g., see
\cite{Bogoyavlensky-CMP1976,OlshanetskyP-PR1981}), but semi-direct
sums of Lie algebras $\bar G$ are normally non-simple (see
\cite{MaXZ-PLA2006}). Our examples in Section 2 are all
non-simple, since the Killing forms on those semi-direct sums of
Lie algebras $\bar G$ are degenerate. However, there still exist
specific non-degenerate bilinear forms on the Lie algebras $\bar
G$, with nice invariance properties, and their corresponding
generalized trace identities, which present Hamiltonian structures
of the enlarged lattice equations.

To conclude, semi-direct sums of Lie algebras provide a good
source of matrix spectral problems for generating integrable
systems, and thus the study of integrable couplings using
semi-direct sums of Lie algebras will enhance our understanding of
classification of integrable systems. We are expecting to see more
research on related topics.

 \vskip 2mm

\noindent{\bf Acknowledgements:} This work was supported in part
by the University of South Florida Internal Awards Program under
Grant No. 1249-936RO and by the National Science Foundation of
China (Grant no. 10471139). Financial support from Dean's office
of the College of Arts and Sciences of the University of South
Florida is also gratefully acknowledged.

%\newpage

\end{document}